\documentstyle[twoside,fleqn,espcrc2]{article}

\newcommand{\be}{\begin{equation}}
\newcommand{\ee}{\end{equation}}
\newcommand{\bea}{\begin{eqnarray}}
\newcommand{\eea}{\end{eqnarray}}
\newcommand{\nn}{\nonumber \\}
\newcommand{\p}[1]{(\ref{#1})}

\def\a{\alpha}\def\b{\beta}

\def\G{\Gamma}
\def\um{{\underline m}}
\def\un{{\underline n}}

\def\bZ {{\bf Z}}
\def\bR {{\bf R}}

\newcommand{\AmS}{{\protect\the\textfont2
  A\kern-.1667em\lower.5ex\hbox{M}\kern-.125emS}}

\title{M(embrane) theory on $T^9$.}

\author{P.K. Townsend\address{DAMTP,
        University of Cambridge \\ 
        Silver St., Cambridge CB3 9EW, U.K.}}

\begin{document}

\begin{abstract}
An `algebraic' approach to M-theory is briefly reviewed, and a proposal is made
for a similar algebraic structure underlying the $T^9$ compactification of 
`M(embrane) theory', i.e. the M(atrix) model with area-preserving
diffeomorphism gauge group.  

\end{abstract}

\maketitle

\section{INTRODUCTION}

This conference provided evidence of a consensus that the most promising 
candidate for a non-perturbative theory of quantum gravity is the conjectured
11-dimensional `M-theory', despite the fact that it has yet to be adequately
defined. That this is possible is due largely to the remarkable power of
supersymmetry, which often allows otherwise unwarranted extrapolation of
semi-classical results. For some time I have been advocating an approach to
M-theory based on the M-theory superalgebra \cite{democracy}, i.e. the `maximal
central extension' of the D=11 supertranslation algebra (originally introduced in
a mathematical context \cite{vans}). This algebra, which is a contraction of
$osp(1|32;\bR)$, is spanned by the 32-component  spinor charge $Q$, the
11-momentum $P$, a two-form charge $Z$ and a 5-form charge $Y$. The only
non-zero (anti)commutator is
\bea 
\label{1}
\{ Q_{\a}, Q_{\b} \} = && (C\G^m)_{\a \b} P_m +
{1\over 2}(C\G^{mn})_{\a \b}Z_{mn} \nn
&&\!\! + {1\over 5!}
(C\G ^{m_1\dots m_5})_{\a \b}Y_{m_1\dots m_5}\, ,
\eea
where $\G^m$ are Dirac matrices, $\G^{m_1\dots m_p}~ (p=2,5)$ their
antisymmetrized products, and $C$ the charge conjugation matrix. In the
Majorana representation spinors are real and $C=\Gamma^0$. Setting
$m=(0,\um)~ (\um=1,\dots, 10)$, the anticommutator \p{1} can be rewritten as
\be
\label{2}
\{ Q, Q\} = P^0(1 + \bar\Gamma)
\ee
where
\bea
\label{3}
P^0\bar\Gamma =&& \big[\G^{0\um}P_\um
+ {1\over2} \G^{0\um\un} Z_{\um\un} \nn
&&\qquad + {1\over 5!}\G^{0\um_1\dots \um_5} Y_{\um_1\dots \um_5}\big]\nn
&&\!\! -\big[ \G^{\um}Z_{0\um} + {1\over 4!} \G^{\um_1\dots\um_4}
Y_{0\um_1\dots\um_4}\big]\, .
\eea 
It is remarkable what one can learn about M-theory by studying this 
superalgebra. For example, the positivity of the anticommutator $\{Q,Q\}$ implies
that $P^0\ge 0$, and that no eigenvalue of $\bar\G^2$ can exceed unity. If
$P^0=0$ we have the D=11 vacuum, or toroidal compactifications of it, in which
supersymmetry is unbroken. Otherwise, the fraction $\nu$ of supersymmetry
preserved is $1/32$ times the number of eigenvectors of $\bar\G$ with eigenvalue
$-1$. Since $\bar\G$ is tracefree, $\nu$ cannot exceed $1/2$. In fact, the
possible values are
\be
\nu = {1\over2}\ ,\ {1\over4}\ ,\ {3\over16}\ ,\ {1\over8}\ ,\ {1\over16}\ ,\
{1\over32}\, .
\ee
The $\nu=1/2$ charge configurations correspond to the `basic' objects of
M-theory{\footnote{This is an oversimplification because there
are additional $\nu=1/2$ `mixed brane' configurations corresponding to
non-marginal bound states of M-2-branes with M-5-branes. However, these can
alternatively be viewed as excited states of the M-5-brane \cite{Dmitri} so there
is a sense in which the `basic' branes are sufficient.}: $P_\um$ is associated
with massless quanta of the effective field theory (or M-waves of D=11
supergravity), $Z_{\um\un}$ is associated with the supermembrane (the M-2-brane)
and $Y_{\um_1\dots \um_5}$ with the superfivebrane (the M-5-brane). The time
components of $Z$ and $Y$ are associated\footnote{This point has been made
independently in \cite{hull}}, respectively, with the D-6-branes and D-8-branes
that arise on compactification to D=10.  
 
Charge configurations with $\nu<1/2$ are associated with intersections of the
`basic' branes. They may also be associated with compactifications on
manifolds of reduced holonomy but, at least in certain limits of moduli space,
these may be viewed locally as intersecting brane configurations. Since
there are some orbifold compactifications that can only be interpreted in this
way (e.g. \cite{wit}), it seems that the intersecting brane perspective is the
more powerful one. In my talk at the conference I described some aspects of work
with Jerome Gauntlett, Gary Gibbons and George Papadopoulos on intersecting 
brane configurations with $\nu=3/16$ \cite{GGPT}, but I intend to review 
this work, and intersecting brane configurations in general, elsewhere. I also
discussed some aspects of how configurations preserving a given fraction of
supersymmetry are related by dualities, but a rather more complete treatment of
this has been provided by others (e.g. \cite{sergio}). Here, I shall take the
opportunity to present some investigations into the algebraic structure
underlying the M(atrix) model conjecture \cite{BFSS} for the microscopic degrees
of M-theory. The conclusion will be that it is remarkably similar to the
M-theory superalgebra just described, which can be viewed as the algebraic
structure underlying the macroscopic semi-classical aspects of M-theory. 

\section{FROM M(ATRIX) to M(EMBRANE)}

The M(atrix) models of M-theory are $SU(k)$ supersymmetric gauge quantum
mechanics (SGQM) models with 16 supersymmetries, of the type first investigated
in \cite{flume}. It was shown in \cite{dHN} that these models can be
viewed as regularizations of the light-cone gauge-fixed D=11 supermembrane
\cite{BST}, which is an SGQM model with the area-preserving diffeomorphism
(APD) group of the membrane as its gauge group. It was observed in \cite{pktb}
that the SGQM model is also the effective action for a condensate of IIA
D-0-branes, and that the continuity of the supermembrane spectrum \cite{dLN} can
therefore be re-interpreted  as a `no force' condition between constituent
D-0-branes. This feature was used in \cite{BFSS} to interpret the first
quantized Hilbert space of the SGQM model as an interacting multi-particle Hilbert space. The existence of a novel $k\rightarrow\infty$ limit was also proposed in which D=11 Lorentz covariance would (hopefully) be recovered 
without the
usual problems associated with removal of a regulator. There is now evidence
(see e.g. \cite{BBPT}) that scattering amplitudes of D=11 supergravity can be
successfully recovered in this approach, with the virtue that the divergences of
quantum D=11 supergravity are now under control, at least in principle. 

For finite $k$ the M(atrix) model Hamiltonian is indeed one in which
the variables are matrices, but this is not true of the $k\rightarrow\infty$
limit. In addition, the restoration of D=11 Lorentz invariance in this limit
appears to be linked to the recovery of the APD group and hence to
the supermembrane interpretation \cite{japanese}. I therefore propose to call
the SGQM model with area-preserving diffeomorphism gauge group the `M(embrane) model'.
It differs from the (first quantized) supermembrane only in the insight derived
from D-branes that M-theory compactified on $T^n$ is described by the
`M(embrane) model on $\tilde T^n$', where $\tilde T^n$ is the dual n-torus to
$T^n$ \cite{Wash,suss}. That is, the SGQM with APD gauge group must be replaced
by an (n+1)-dimensional gauge field theory with APD gauge group, where the
n-dimensional space is $\tilde T^n$. According to this prescription, M-theory 
on $S^1$, alias (second quantized) IIA superstring theory, is described by a
(1+1)-dimensional APD gauge theory. This has recently been verified in
considerable detail \cite{Motl,DVV}. 

For $n>1$ we should find the appropriate
U-duality group of toroidally-compactified M-theory \cite{CJ,HT}. For $n=2$
this group is $Sl(2;\bZ)$, which indeed arises in the M(atrix)
models as the modular group of $\tilde T^2$. More generally, a gauge theory
on $\tilde T^n$ has a manifest invariance under the $Sl(n;\bZ)$ modular group
of $\tilde T^n$, but this is a proper subgroup of the U-duality group when $n>2$. To recover the full U-duality group for $n\ge3$ one must include the magnetic degrees of freedom of M(atrix) models, corresponding to 5-branes in
M-theory. At this point we can see that the M(atrix) model approach is an `optimally democratic' formulation of M-theory, in the sense of \cite{democracy}. That is, it incorporates into a new type of perturbation theory all of the {\sl electric} degrees of freedom of M-theory (as it was argued in \cite{democracy} that a supermembrane theory might do). Since the supermembrane is the strong coupling limit of the IIA superstring, the M(atrix) model approach is non-perturbative with respect to superstring theory, but it is still perturbative as a formulation of M-theory. A truly non-perturbative, and fully democratic, formulation of M-theory must incorporate the magnetic degrees of freedom. 

For $n=3$ it is clear how this is to be done \cite{suss}. The U-duality group
of $T^3$-compactified M-theory is $Sl(3;\bZ)\times Sl(2;\bZ)$, the $Sl(2;\bZ)$
factor being the `electromagnetic' duality group acting on dyonic membranes
\cite{izq} (equivalently, wrapped 2-brane/5-brane bound states in D=11). The
M(atrix) models on $\tilde T^3$ have an obvious $Sl(3;\bZ)$
invariance but they are also conjectured to be invariant under a
non-perturbative $Sl(2;\bZ)$ electromagnetic S-duality \cite{sen}.
For $n\ge3$ the situation is further complicated by the non-renormalizability of
(n+1)-dimensional gauge theories. The M-theory dynamics should be governed by
a superconformal fixed point theory to which the gauge theory flows in the
ultra-violet, but this UV fixed-point theory is certainly not defined by the
gauge theory itself. For $n=4$ the M(atrix) model is nominally a 
(4+1)-dimensional gauge theory with a manifest $Sl(4;\bZ)$ invariance but the
instantonic solitons of D=5 gauge theories can be interpreted as the
Kaluza-Klein modes in an extra dimension, so that when these are taken into
account the $Sl(4;\bZ)$ invariance is enhanced to the full U-duality group
$Sl(5;\bZ)$ \cite{kk}. This theory can be viewed as the compactification on
$S^1$ of a non-local (2,0) superconformal theory in D=6, invariant under the D=6
U-duality group $SO(5,5)$; this D=6 theory governs the dynamics of M-theory on
$T^5$ \cite{Seiberg}. The gauge theory instantons are now (non-critical)
strings. 

In general, one expects the (n+1)-dimensional gauge theory governing M-theory on $T^n$ to be a
compactification of the corresponding (n+2)-dimensional gauge theory governing
M-theory on $T^{n+1}$, so the superconformal fixed point of (5+1)-dimensional
gauge theories is effectively the `master' theory for all $T^n$
compactifications of M-theory with $n\le5$.  There is no reason to stop at 
$n=5$, however. Recent candidates involving membranes rather than strings have 
been proposed for the theory governing M-theory on $T^6$ \cite{Moore},
and the issue of how U-duality is realized for $n\ge6$ has been studied
\cite{elitzur}. A feature of the $n\le8$ cases is that the realization of
U-duality does not depend on taking the $k\rightarrow\infty$ limit of the
$SU(k)$ gauge theory. In other words, U-duality does not distinguish between
M(atrix) and M(embrane). This seems unlikely to remain true for M-theory
on $T^9$ because the conjectured U-duality group in this case is a
discretization of the infinite-dimensional group with Lie-algebra $E_9$
\cite{Julia,HT}. Rather, it seems likely that one will need the
infinite-dimensional APD group. This is supported by some suggestions in
\cite{emil} in which an APD D=10 gauge theory is extracted from the N=(2,1)
heterotic string approach to M-theory. Local gauge theories are anomalous in
D=10 but chiral anomalies can be cancelled by non-local counterterms, which is
presumably admissable in a non-local theory.  

These works point to a non-local D=10 superconformal UV-fixed point theory
having some connection to area-preserving diffeomorphisms as the true `master'
theory underlying the M(atrix) model approach. In any case, the existence of 
some such theory, which I will call `M(embrane) theory', is clearly required by
the conjecture of \cite{BFSS}. The problem is that it is apparently
excluded by the absence of a superconformal group; according to the standard
classification \cite{nahm}, superconformal groups exist for $D\le6$ but not for
$D>6$. Here I will suggest a resolution of this puzzle.

\section{M(EMBRANE) SUPERALGEBRA}

We first note that the classification of superconformal groups in \cite{nahm} is
based on the Coleman-Mandula theorem, which requires the bosonic symmetry group
to be the direct product of the spacetime symmetry group (the D=10 conformal
group $SO(10,2)$ in the case of interest here) and some internal symmetry group.
The Coleman-Mandula theorem holds for {\it local} QFT. Once locality is
abandoned there is no obstacle to superconformal invariance for D=10. In fact,
there is then a  natural candidate for the D=10 superconformal group. 

As motivation, let us first recall that D=10 gauge theories have instantonic
fivebranes \cite{pktc} which (as in D=4 \cite{olive}) lead to a central 
extension of the D=10 N=1 supertranslation algebra spanned by the 16
(independent) component chiral spinor charges $Q$, the 10-momentum $P$ and a
self-dual 5-form charge $Z^+$. The only non-zero (anti)commutator is
\bea
\{ Q_\alpha, Q_\beta\} =&& (C\Gamma^m{\cal P}^+)_{\alpha\beta} P_m \nn
&& +  {1\over 5!} (C\Gamma^{mnpqr})_{\alpha\beta} Z^+_{mnpqr}
\label{4}
\eea
where ${\cal P}^+$ is the positive chirality projection operator on spinors.
As noted in \cite{democracy}, this is the `maximal central extension' of the N=1
D=10 supertranslation algebra. On the basis of this algebra one would expect the
stress-tensor supermultiplet to include a 6-form current associated to the
5-form charge. In fact\footnote{I thank Paul Howe for reminding me of this.},
there is a non-local 256-component superconformal current multiplet in D=10 for
which the 128 bosonic components consist of the traceless stress tensor $T_{mn}$
and a 6-form current $J_{mnpqrs}$ \cite{berg}. 

This confirms both that we must abandon locality and that, in doing so, we should
seek a superconformal extension of \p{4}, which must involve an antichiral
conformal supersymmetry charge $S^\alpha$ (we use the position of indices to
keep track of chirality). It must also contain both the D=10 conformal algebra
$so(10,2)$ and the supertranslation algebra \p{4} as sub-(super)algebras. It is
not clear to me whether the solution to these requirements is unique but there
is an obvious solution: the superalgebra $osp(1|32;\bR)$. The additional
anticommutation relations are
\bea
\{ S^\alpha, S^\beta\} &=& (C\Gamma^m {\cal P}^-)^{\alpha\beta} K_m \nn
&& + (C\Gamma^{mnpqr})^{\alpha\beta} Z^-_{mnpqr} \label{5}\\
\{S^\alpha, Q_\beta\} &=& \delta^\alpha_\beta D + {1\over2}
(\Gamma^{mn})^\alpha{}_\beta M_{mn} \nn
&&+ {1\over 4!}(\Gamma^{mnpq})^\alpha{}_\beta Y_{mnpq} 
\label{6}
\eea
where $K,D,M$ are the generators of conformal boosts, dilations and Lorentz
transformations, respectively. The anti-self-dual 5-form charge $Z^-$ is the
conformal analogue of $Z^+$ while $Y$ is a new 4-form charge. 

The two spinor charges $Q$ and $S$ can be assembled into a 32-component real
spinor of $SO(10,2)$ with components ${\cal Q}_A = (Q_\alpha, S^\beta)$.
Introducing the $SO(10,2)$ Dirac matrices $\Gamma^M$ and their antisymmetrized
products, $\Gamma^{MN}$ etc., we can rewrite the anticommutators of \p{4},
\p{5} and \p{6} in the manifestly $SO(10,2)$-covariant form
\bea
\{ {\cal Q}_A, {\cal Q}_B\}&& = (C\Gamma^{MN}{\cal P}^+)_{AB} L_{MN}\nn
&& \!\!\! + {1\over 6!} (C\Gamma^{MNPQRS})_{AB} T^+_{MNPQRS}
\label{7}
\eea
where $C$ is now the $SO(10,2)$ `charge conjugation' matrix, and ${\cal P}^+$ the
chiral projector on $SO(10,2)$ spinors. $L$ comprises the generators $(K,D,M,P)$
of $SO(10,2)$, while $T^+$ (composed of $Z^\pm$ and $Y$) is a self-dual antisymmetric 6th rank tensor of
$SO(10,2)$. In fact, the algebra is $Sp(32)$ covariant with $(L,T^+)$ in the
${\bf 528}$ adjoint representation and ${\cal Q}$ in the ${\bf 32}$
representation. Similar algebras have appeared elsewhere in the context of
speculations concerning a 12-dimensional extension of M-theory \cite{bars}, but
here there is an important difference; the charge $T^+$ does {\it not} commute
with $L$.  The commutation relations of $(L,T^+)$ are those of $sp(32)$ for which
the  matrices $\Gamma^{MN}{\cal P}^+$ and $\Gamma^{MNPQRS}{\cal P}^+$ form the
32-dimensional representation. The full set of (anti)commutation relations are
those of $osp(1|32;\bR)$. 

\section{COMMENTS}

I began this article by explaining how the `M-theory superalgebra', a
contraction of $osp(1|32;\bR)$, encapsulates much of what we can learn from a
semi-classical analysis of M-theory. I then turned to the consideration of
the M(atrix), or M(embrane), model approach to the microscopic theory and
asked whether there were some comparable algebraic structure underlying it.
My conclusion was that there is and that it is $osp(1|32;\bR)$. As far as I
can see this is just a coincidence; the algebras are actually quite different.
The M-theory algebra does not include the Lorentz group (this has to be
considered separately as its automorphism group) and the M(embrane) algebra is
superconformal. Nevertheless, the possibility of a deeper connection
hoefully provides a justification for the joint discussion here of both
algebraic structures.  

Of course, identifying the superconformal group associated with M(embrane) 
theory is only a small step towards its construction. Other clues are the
expected $E_9$ invariance, and the fact that it should involve 5-branes. 
Also, it should reduce, on $T^4$ compactification to the (2,0) D=6 
superconformal theory underlying M(atrix) theory on $\tilde T^5$; the  strings
of the latter model are presumably $T^4$-wrapped 5-branes of  M(embrane) theory.
Similarly, the membranes of the theory proposed to govern the $T^6$
compactification of M-theory are presumably $T^3$-wrapped 5-branes of
M(embrane) theory.

\end{document}